\documentclass[reprint, amsmath, amssymbol, twocolumn, prb]{revtex4-2}

\usepackage[colorlinks = true, citecolor = blue, urlcolor = blue, linkcolor = black]{hyperref}
\usepackage{graphicx}
\usepackage{dcolumn}
\usepackage{bm}
\usepackage{siunitx}
\usepackage{float}

\begin{document}


\title[Sinter Paper]{Effect of boundary condition on Kapitza resistance between superfluid \textsuperscript{3}He-B and sintered metal}

\author{S.~Autti}
\author{A.\,M.~Gu\'{e}nault}
\altaffiliation{Deceased}
\author{A.~Jennings}
\email{a.jennings1@lancaster.ac.uk}
\author{R.\,P.~Haley}
\author{G.\,R.~Pickett}
\author{R.~Schanen}
\author{V.~Tsepelin}
\author{J.~Vonka}
\altaffiliation{Current address: Paul Scherrer Institute, Forschungsstrasse 111, 5232 Villigen PSI, Switzerland}
\author{D.\,E.~Zmeev}

\affiliation{Department of Physics, Lancaster University, Lancaster, LA1 4YB, United Kingdom}

\author{A.\,A.~Soldatov}
\affiliation{P.\,L. Kapitza Institute for Physical Problems of RAS, 119334 Moscow, Russia}

\date{\today}

\begin{abstract}
Understanding the temperature dependence of thermal boundary resistance, or Kapitza resistance, between liquid helium and sintered metal has posed a problem in low temperature physics for decades. In the ballistic regime of superfluid \textsuperscript{3}He-B, we find the Kapitza resistance can be described via scattering of thermal excitations (quasiparticles) with a macroscopic geometric area, rather than the sintered metal's microscopic area. We estimate that a quasiparticle needs on the order of 1000 collisions to successfully thermalize with the sinter. Finally, we find that the Kapitza resistance is approximately doubled with the addition of two mono-layers of solid \textsuperscript{4}He on the sinter surface, which we attribute to an extra magnetic channel of heat transfer being closed as the non-magnetic solid \textsuperscript{4}He replaces the magnetic solid \textsuperscript{3}He.
\end{abstract}

\pacs{Valid PACS appear here}
\maketitle

%

\section{\label{Intro} Introduction}

The thermal boundary resistance, or Kapitza resistance, $R_{\mathrm{K}}$ between liquid helium and sintered metal is a common limiting factor in the pursuit of ultra-low temperatures. As temperature decreases the thermal boundary resistance $R_{\mathrm{K}}$ increases, reducing the cooling power attainable in dilution refrigerators. According to acoustic mismatch theory, phonon-phonon transfer should yield a boundary resistance proportional to inverse temperature cubed, $R_{\mathrm{K}} \propto T^{-3}$ \cite{Nakayama1989}. A common way of decreasing the Kapitza resistance between liquid helium and metal and increasing refrigerator performance is to increase the surface area using sintered metal powders. However, the problem of Kapitza resistance between sintered metal and liquid \textsuperscript{3}He is still yet poorly understood despite the large scale use of dilution refrigerators, in part due to sintered metals' complicated geometry \cite{Nakayama1989}. At very low temperatures, $T\leq \SI{10}{\milli\kelvin}$, the boundary resistance deviates significantly from the prediction of acoustic mismatch theory. A lower than expected Kapitza resistance is observed in general, and with a temperature dependence of $R_{\mathrm{K}} \propto T^{-1}$ \cite{Franco1984}. In superfluid \textsuperscript{3}He-B at ultra-low temperatures (about $0.3~T_c$ or \SI{0.3}{\milli\kelvin} at zero pressure) the mean free path of the thermal excitations exceeds the cell dimensions and hydrodynamic picture becomes inadequate. In this ballistic regime previous measurements reported an exponential dependence \cite{Castelijns1985, Parpia1985} as well as quadratic or cubic relationships \cite{Voncken1996} of Kapitza resistance with temperature.

It is important to also consider the possible surface effects of solid helium at low temperatures as a few layers of solid helium will form on the surfaces due to van der Waal's forces \cite{Tholen1993, Levitin2013}. It is thought that the magnetic properties of \textsuperscript{3}He could play a large role in the smaller than expected Kapitza resistance, by the creation of a magnetic channel of energy exchange with magnetic impurities in the sintered metal \cite{Koenig1998, Koenig1998a}.

Recently, it has become a popular technique in ultra-low temperature physics to pre-plate the surfaces of the experimental cell and objects with solid \textsuperscript{4}He. \textsuperscript{4}He preferentially adsorbs to surfaces over \textsuperscript{3}He due to its lower zero-point energy. Control over the number of \textsuperscript{4}He layers allows fine tuning of the surface scattering specularity. This technique has been used for the stabilization of the newly discovered polar phase in nematic aerogels \cite{Dmitriev2015} and the suppression of the B phase in thin slab geometries when the scattering at the slab surface becomes completely specular \cite{Levitin2013, Tholen1993}. Importantly, solid \textsuperscript{4}He is also non-magnetic. In the ballistic regime of the B phase, heat transfer should be dominated by quasiparticle collisions with the surface. The difference in magnetic and non-magnetic quasiparticle scattering has been found to have a significant effect on the stabilisation of the A phase \cite{Zimmerman2020, Heikkinen2019} and of the polar phase \cite{Dmitriev2018} in anisotropic media. It is therefore likely  that the magnetic properties of solid \textsuperscript{3}He are an important factor in the Kapitza resistance. Hence  pre-plating surfaces with non-magnetic \textsuperscript{4}He should  change the effective Kapitza resistance, a key consideration for future experiments at ultra-low temperatures.

In this work we demonstrate the difference in Kapitza resistance between superfluid \textsuperscript{3}He-B and sintered silver with and without pre-plating of the surface with 2 layers of \textsuperscript{4}He. We compare the measured Kapitza resistance to earlier work \cite{Castelijns1985, Carney1989} and find very good agreement with the reported exponential temperature dependence. Our observed difference in Kapitza resistance between solid \textsuperscript{3}He and solid \textsuperscript{4}He coverage provides evidence for the role of the magnetic channel in decreasing the Kapitza resistance between metal and liquid \textsuperscript{3}He. We use a simple model to calculate the increase in probability of a quasiparticle collision with sintered metal eventually resulting in quasiparticle recombination. We find that sinter covered with \textsuperscript{3}He offers about double the probability of recombination.

We have analysed the results of three experiments and conclude that in the ballistic regime it is the geometric area of the sinters that needs to be considered in the calculation of the Kapitza resistance. We define the geometric area as the area of plates with direct contact to the liquid in the experimental volume which provides consistent results between measurements. The total (interfacial) area of all sintered plates or the area calculated from the sponge-like sinter geometry at the microscopic level do not provide consistent results between different measurements. Similar conclusions have been made about the sinter surface area in contact with a saturated $^3$He-$^4$He solution \cite{Cousins1994}. This makes intuitive sense, as a quasiparticle created in the volume has little chance of colliding with sintered plates that have other plates between them and the quasiparticle.

\section{Experimental Details}

\begin{figure}[t]
\includegraphics[width=\linewidth]{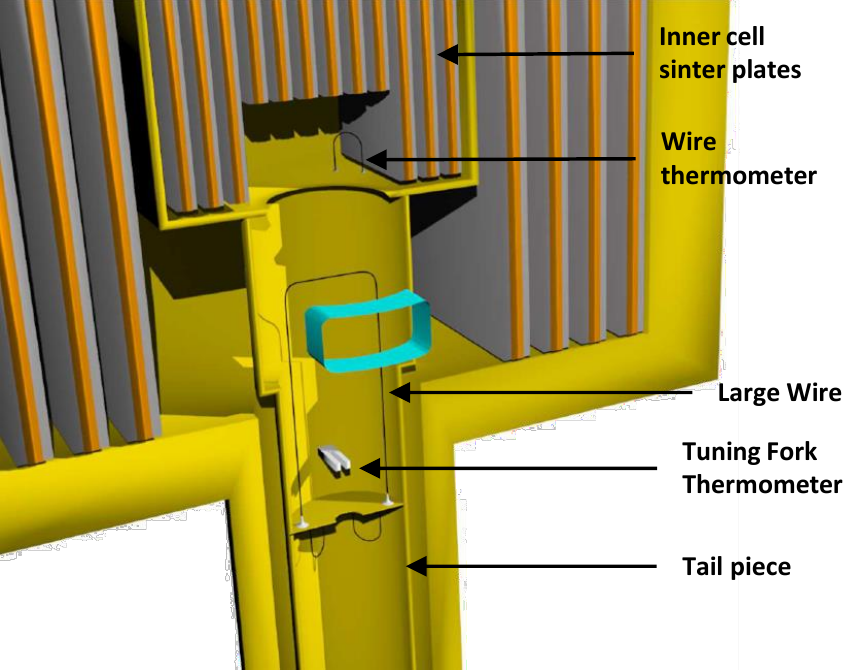}
  \caption{\label{fig:cell} The experimental cell. The inner cell contains eighty \SI{0.2}{\milli\meter} thick and eight \SI{1.1}{\milli\meter} thick copper plates with dimensions \SI{49}{\milli\meter} by \SI{28}{\milli\meter}. The plates are sintered with a silver powder, giving a total microscopic sinter area of \SI{80}{\meter\squared}. A \SI{2}{\centi\meter} by \SI{2}{\centi\meter} by \SI{1}{\centi\meter} geometric volume is formed by cutting the middle plates. The cuts from these plates are in an alternating pattern of a \SI{1.9}{\centi\meter} by \SI{1.1}{\centi\meter} rectangle followed by a \SI{2.0}{\centi\meter} by \SI{1.0}{\centi\meter} rectangle to create \SI{1}{\milli\meter} protrusions that increase the surface area without changing the volume. The geometric volume contains the vibrating loop wire thermometer. A cylindrical volume containing the large wire and quartz tuning fork is connected via a circular cutout to the bottom of the geometric volume. The cylindrical volume also has a tail piece separated by a small hole.}
\end{figure}

Our experimental apparatus is a Lancaster-style nested nuclear demagnetization cell attached to a custom-built dilution refrigerator (shown in Fig.~\ref{fig:cell}). The inner cell contains 8 copper plates of thickness \SI{1.1}{\milli\meter} and 80 copper plates of thickness \SI{0.2}{\milli\meter} which are used as a refrigerant for nuclear demagnetization. A silver powder of \SI{70}{\nano\meter} particle size is sintered to both sides of each copper plate with a filling factor of 0.5 and results in a microscopic surface area of \SI{80}{\meter\squared} \cite{Keith1984}. From the total surface area, which is more or less equal to the total microscopic sinter surface area, we can calculate the amount of gas needed to pre-plate all surfaces with 2 layers of solid \textsuperscript{4}He \cite{Autti2020}. This is accomplished by inserting a small amount of \textsuperscript{4}He gas before filling the cell with \textsuperscript{3}He to saturated vapour pressure. After demagnetizing to the required magnetic field (or temperature) the copper plates act as a thermal bath. 

Inside the cell we have a set of mechanical probes: A relatively large moving wire with a diameter of \SI{135}{\micro\meter}, a small vibrating loop wire with a diameter of \SI{4.5}{\micro\meter} and a small quartz tuning fork. The damping in ballistic \textsuperscript{3}He-B measured via mechanical resonance widths $\Delta f$ of the vibrating loop and tuning fork is determined by the number of quasiparticle collisions, and hence the number density of quasiparticles in the superfluid. In the ballistic regime the quasiparticle density is exponential with temperature, thus vibrating resonators can be very sensitive thermometers \cite{Guenault1986}. The large wire can be moved at constant velocity with a direct current in a magnetic field which generates a Laplace force, which we term a DC pulse \cite{Zmeev2013}. Moving the wire heats the experimental cell and increases the number of quasiparticles above the thermal equilibrium density \cite{Bradley2016}. Hence monitoring the width response of the vibrating loop and tuning fork during a DC pulse allows us to detect the heating and subsequent cooldown. Note, that the actual temperature change is small due to the exponential dependence of the resonance width on temperature. 

\section{Results}

Initially, the bulk quasiparticle population is in thermal equilibrium with the thermal bath and the thermometer is resonating with a base resonance width of $\Delta f_{base}$.

\begin{figure}[t]
  \includegraphics[width=\linewidth]{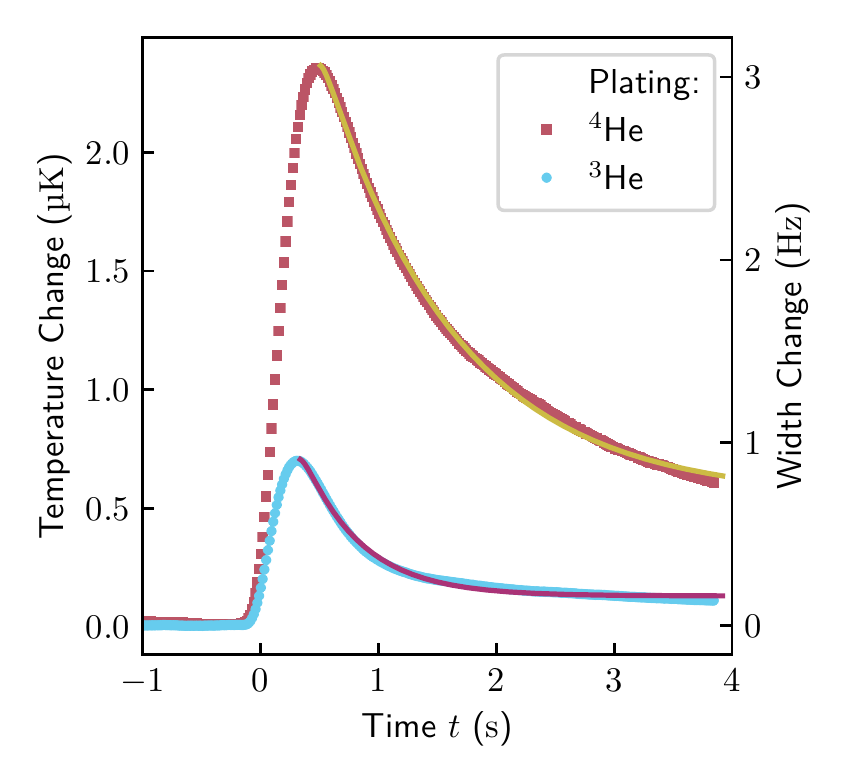}
  \caption{\label{fig:resp} Comparison of the response of the vibrating loop wire with diameter \SI{4.5}{\micro\meter} for pure \textsuperscript{3}He and two layers of solid \textsuperscript{4}He coverage on the cell surface at temperatures of \SI{197}{\micro\kelvin} ($\Delta f_{base} = \SI{30.4}{\hertz}$) and \SI{193}{\micro\kelvin} ($\Delta f_{base} = \SI{25.9}{\hertz}$), respectively. The resonance width is related to temperature by the equation $\Delta f = \Delta f_{\mathrm{vac}} + \gamma \exp\left(-\Delta_B/k_{\mathrm{B}}T\right)$, where  $\Delta f_{\mathrm{vac}} = \SI{92}{\milli\hertz}$ is the resonance width in vacuum and $\gamma$ is a fitting constant~\cite{Baeuerle1998}. The corresponding width change for the \textsuperscript{4}He data is given by the right y axis. The \textsuperscript{3}He data was measured at a slightly different temperature, and hence the width change is $5\%$ larger for a given temperature change. Solid lines represent the fits from Eq.~(\ref{eq1:response}).}
\end{figure}

After a DC pulse is started, the released heat goes directly into the liquid and increases the bulk quasiparticle population, detected by a rapid increase in thermometer width up to a peak value. The quasiparticles excited begin colliding with surfaces where the area is dominated by the silver sinter. In the ballistic regime these collisions are either near-elastic or result in the quasiparticles losing enough energy to successfully recombine as a Cooper pair. We expect an exponential decay of the quasiparticle density until thermal equilibrium is reached with the surroundings, measured by the thermometer width reaching the base value. Figure \ref{fig:resp} shows a typical response of a thermometer width to a DC pulse that we model as \cite{Winkelmann2007}

\begin{equation}
\Delta f = \Delta f_{base} + H \frac{\tau_b}{\tau_b+\tau_w}\left( e^{-t/\tau_b} -e^{-t/\tau_w}\right)\mathrm{.}
\label{eq1:response}
\end{equation}

\noindent Here $H$ is a constant describing the amplitude  of the width response and $\tau_w$ is the response time of the thermometer, determined by the resonance width and is approximately equal to $1/\left(\pi\Delta f_{base}\right)$. The decay time constant $\tau_b$ is governed by the effective boundary resistance $R_\mathrm{K}$, area $A$ and the heat capacity $C_B$ of superfluid \textsuperscript{3}He-B \cite{Parpia1985}

\begin{equation}
A R_{\mathrm{K}} = \frac{\tau_b}{C_{B}}\mathrm{.}
\label{eq2:Kapitza}
\end{equation}

The heat capacity of \textsuperscript{3}He-B in the ballistic regime is determined by the number of quasiparticles and exponentially depends on temperature $C_B \propto \exp\left({-\Delta_B / k_\mathrm{B} T}\right)$, where $\Delta_B$ is the superfluid energy gap of \textsuperscript{3}He-B and $k_\mathrm{B}$ is the Boltzmann constant. 

\begin{figure}[t]
  \includegraphics[width=\linewidth]{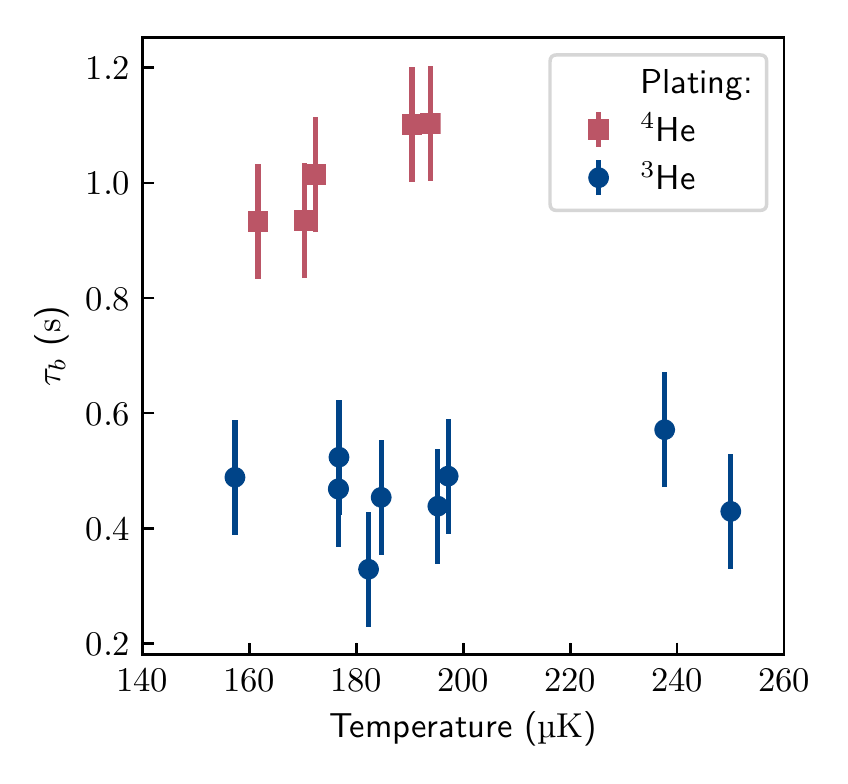}
  \caption{\label{fig:tau_B} Comparison of quasiparticle relaxation time constant $\tau_b$ as measured by the vibrating loop thermometer. The average $\tau_b$ is $0.5~ \pm$~\SI{0.1}{\second} and $1.0 \pm\SI{0.1}{\second}$ for solid \textsuperscript{3}He and solid \textsuperscript{4}He coverage, respectively. Absence of discernible variation over temperature is consistent with an exponential dependence of $R_{\mathrm{K}}$.}
\end{figure}

Figure~\ref{fig:resp} shows the different responses of the vibrating loop thermometer to identical pulses for pure \textsuperscript{3}He and \textsuperscript{4}He pre-plated cases. The decay constant $\tau_b$ extracted for \textsuperscript{4}He pre-plated surfaces is $1.0 \pm\SI{0.1}{\second}$ which is approximately double the $0.5~ \pm$~\SI{0.1}{\second} constant observed in pure \textsuperscript{3}He. Similar results were obtained for the quartz tuning fork. Interestingly, the time constant was found to not vary with the wire velocity of the DC pulse. The time constant is thus independent of the energy released by the DC pulse at least when the energy released is varied from \SI{5}{\pico\joule} to \SI{100}{\pico\joule} (wire velocities of \SIrange{2}{50}{\milli\meter\per\second}). Furthermore, the thermalization time constant $\tau_b$ is independent of temperature for both types of coverages, which is demonstrated in Fig~\ref{fig:tau_B}. According to Eq.~(\ref{eq2:Kapitza}), this demonstrates an exponential temperature dependence on Kapitza resistance in both cases, and leads to the conclusion that pre-plating with solid \textsuperscript{4}He doubles the Kapitza resistance.

In the above analysis we have assumed that only the bulk \textsuperscript{3}He heat capacity is relevant. It is known that the solid \textsuperscript{3}He layer in a magnetic field and at low temperatures can have a large contribution. Elbs \textit{et al.} measured the surface contribution to the overall heat capacity for a \SI{0.13}{\centi\meter\cubed} container of \textsuperscript{3}He and found that at higher temperatures around \SI{0.16}{\milli\kelvin} to \SI{0.19}{\milli\kelvin} in similar magnetic field to our measurements the heat capacity of solid and liquid helium were roughly equal \cite{Elbs2007}. For our much larger volume of \SI{8.6}{\centi\meter\cubed} the contribution to heat capacity from the superfluid \textsuperscript{3}He-B is much more than the contribution to heat capacity from any solid layers, thus we neglect this heat capacity. 

\begin{table*}
    \centering
\begin{ruledtabular}
\begin{tabular}{lccccr}
Cell &  Experimental Volume (\SI{}{\centi\meter\cubed}) & \multicolumn{3}{c}{Sinter Area} & Sinter Mass (\SI{}{\gram}) \\
\hline \\
& & Interfacial (\SI{}{\meter\squared}) & Microscopic (\SI{}{\meter\squared}) & Geometric (\SI{}{\centi\meter\squared}) & \\
\hline\\
This work  & 8.6 & 0.21 & 80 & 36 & 96 \\
Castelijns \textit{et al.}~\cite{Castelijns1985}  & 1.0 & 0.011 & 22 & 3.0 & 28 \\
Carney \textit{et al.}~\cite{Carney1989} & 1.0 & 0.17 & 41 & 3.0 & 49 \\
\end{tabular}
\end{ruledtabular}
\caption{\label{tab:cells} Experimental cells for each measurement. The interfacial area is the area of all plate faces covered in sinter and microscopic area is the area of the sintered powder's sponge-like surface. Geometric area is the sintered plate surfaces that face the experimental volume containing the large wire with no other sintered surfaces between it and the volume (see text). The experimental volume quoted for this work excludes the tail piece, in which a small hole limited heat flow into the tail piece volume. However, the tail piece volume is not large and its inclusion would not change the results significantly.}
\end{table*}

\section{Discussion}

The exponential temperature dependence is entirely consistent with previously reported by Parpia \cite{Parpia1985}, Castelijns \textit{et al.} and Carney \textit{et al.} \cite{Castelijns1985, Carney1989}. Thermalization time $\tau_b$ reported by Castelijns \textit{et al.}, is about ten times larger. In that work an electrical heater was used to create heating bursts, which likely created a bubble of normal fluid around the heater and altered the time constant \cite{Castelijns1985}. The advantage of our method is the direct creation and relaxation of quasiparticles. The wire moving at supercritical velocities does not destroy the bulk superfluid state \cite{Bradley2016}. Therefore, we observe purely the dynamics of the quasiparticles, which in the ballistic regime governs the heat transport properties of the superfluid.

Consider a superfluid quasiparticle excited by the DC pulse. The quasiparticle travels with group velocity $v_g$ which is approximately \SI{20}{\meter\per\second} at these temperatures and pressure. Due to the superfluid energy gap, when the quasiparticle scatters with a wall it has a chance to either scatter elastically or lose energy $\Delta_B$ and recombine into a Cooper pair. The recombination results in a reduction of the overall superfluid temperature. However, losing enough energy for recombination after scattering with cell walls is extremely unlikely and quasiparticle-quasiparticle interactions can be neglected in the ballistic regime, hence only scattering with sinter matters.  The pore size in the sinter is roughly equal to the coherence length of the superfluid pairs which is around \SI{80}{\nano\meter} at zero pressure \cite{Serene1983, Franco1984}. Hence, the sinter could appear as a wall of metal and high normal fluid density to any approaching quasiparticle \cite{Hall1992}.

We can estimate the number of collisions needed for a quasiparticle to eventually recombine by considering its motion in a cubic box of dimension $d$ with only one thermally conductive wall due to the sintered copper plates. The time it takes to traverse from one wall to another is roughly $t_{col} = d/v_g$ and the average time taken for there to be no more collisions is approximately $\tau_b$. Therefore the probability of collision successfully leading to recombination is $P = \frac{t_{col}}{6\tau_{b}}$. This gives an estimate on the order of 1000 collisions needed for a quasiparticle to lose its energy with the sinter and recombine.

For \textsuperscript{4}He coverage we also observed a similar magnetic field dependence as measured by Osheroff and Richardson \cite{Osheroff1985} with pure \textsuperscript{3}He. About a 20\% increase was observed in $\tau_b$ as the field was doubled from \SI{63}{\milli\tesla} to \SI{125}{\milli\tesla}. This confirms the conclusion that magnetic impurities in the sinter also reduce the Kapitza resistance, and that this reduction is independent of the magnetic layers of solid $^3$He that form on all surfaces in the absence of $^4$He preplating.

The simple quasiparticle collision model described above assumes that the sinter is a flat surface in just one side of the cell. In contrast, our experimental cell has many plates of sinter and many of these plates are behind other plates. However, we may assume that a quasiparticle in the experimental volume has a significant chance of scattering only with sinter that is near the experimental volume and has no other plates or walls between it and the volume. We define a ``geometric area" which consists of only the faces of the sintered plates that are in direct contact with this volume. We treat this area as if it has no gaps, meaning a quasiparticle cannot travel around the smalls gap between plates and the cell walls. This also means we assume the areas made of a collection of plates pushed together, such as the one just above the vibrating wire thermometer, as continuous. In reality there are small gaps. We can then test our assumption and the model by comparing our values of thermal boundary resistance to those obtained by Castelijns \textit{et al.} and Carney \textit{et al.} using the same silver powder and sintering technique as used in this work \cite{Keith1984}. 

We fit zero pressure data from Castelijns \textit{et al.} and Carney \textit{et al.} with a function that uses the BCS gap $\Delta_B = 1.76 k_B T_c$. We compare several different areas for use in Eq.~(\ref{eq2:Kapitza}): The geometric area described above; the microscopic area, the area of all the sinter when not treated as flat surface at the microscopic level; and the interfacial area, calculated by the area of all plates in the cell. Table~\ref{tab:cells} gives details of the volumes and different areas for each experiment.

\begin{figure}[t]
  \includegraphics[width=\linewidth]{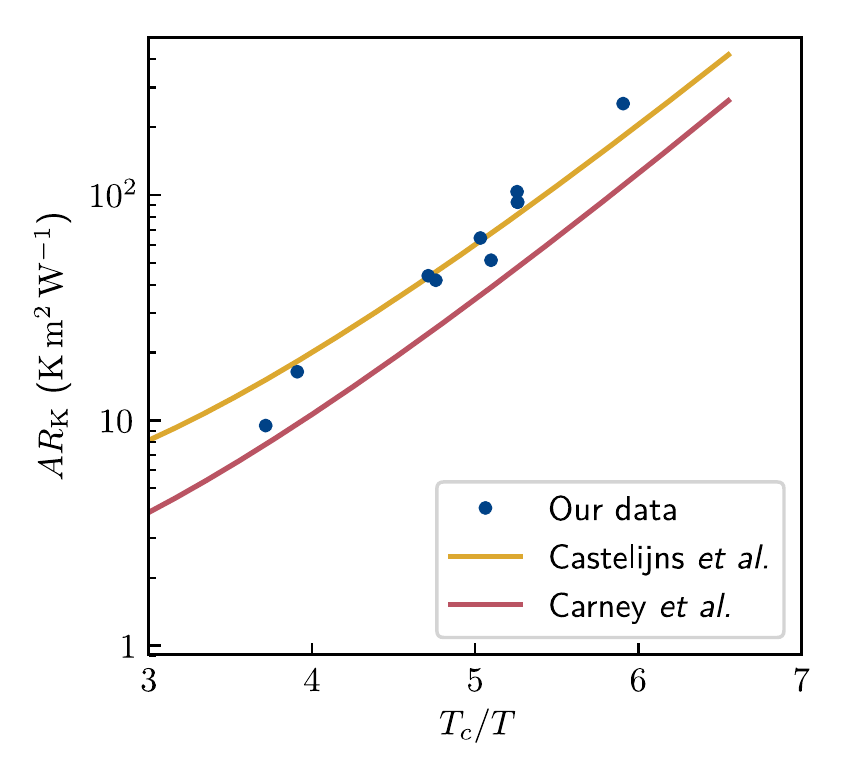}
  \caption{\label{fig:Resistance} The effective thermal boundary resistance $R_{\mathrm{K}}$ multiplied by the ``geometric" sinter scattering area $A$ for quasiparticles created by the wire as a function of inverse temperature. $T_c=$\,\SI{929}{\micro\kelvin} is the critical temperature of superfluid \textsuperscript{3}He at zero pressure. The gold and red lines are fits for data from Refs. \cite{Castelijns1985} and \cite{Carney1989}, respectively, and converted into thermal boundary resistance. For comparison, our data shown is only in pure \textsuperscript{3}He. They lie mostly on the gold line, or between the two lines if the tailpiece volume is included.}
\end{figure}

We find that the geometric area gives good agreement with our results, shown in Fig.~\ref{fig:Resistance}. Moreover, this is the only area that gives any agreement. Notably, in both experiments by Carney \textit{et al.} and Castelijns \textit{et al.} have the exact same cell geometry in terms of geometric area, but the ratio between the measured boundary resistances at the lowest temperatures was about 2.5. Our data points lie closer to the results of Castelijns \textit{et al.}, or between the two lines if the tail piece volume is included in the calculation of the heat capacity in Eq.~(\ref{eq2:Kapitza}). 

The difference between the results of Castelijns \textit{et al.} and Carney \textit{et al.} comes from the assumption that gaps between plates do not matter. In reality, a quasiparticle may go between the gaps and begin colliding within the plates. Since there is a much shorter distance between these two close plates, a quasiparticle has much less time between the collisions resulting in quicker recombination. Naturally, the chances of a quasiparticle going inside a gap will increase with the number of gaps. In Castelijns \textit{et al.} the cell had twelve \SI{1}{\milli\meter} thick plates, and a square was cut out of seven to form a \SI{1}{\centi\meter\cubed} box. In Carney \textit{et al.}, ninety-two plates of \SI{0.1}{\milli\meter} were used, resulting in many more gaps and the lower Kapitza resistance seen in Figure~\ref{fig:Resistance}. In our cell, there is a combination of both thick and thin plates, and our values lie closer to those reported from Castelijns \textit{et al.}. Therefore, the sinter plate layout is the likely explanation for the different outcomes obtained by these three experiments.

The difference between the two types of coverage, we believe, is due to different magnetic properties of the two. It has long been thought that a magnetic channel for transfer of heat is key to explaining the observed difference in Kapitza resistance between experiment and theory \cite{Avenel1973}. Further evidence was added by K\"onig \textit{et al.}, who saw a large difference in the performance of different brands of silver powder \cite{Koenig1997, Koenig1998, Koenig1998a}. They concluded that Ulvac powder has a larger content of magnetic impurities and a lower Kapitza resistance at millikelvin temperatures. Incidentally, we, Osheroff and Richardson \cite{Osheroff1985}, Castelijns \textit{et al.}~\cite{Castelijns1985} and Carney \textit{et al.} \cite{Carney1989} all used the \SI{70}{\nano\meter} Ulvac powder, which has the highest magnetic impurity content of all brands and sizes measured by K\"onig \textit{et al}.  

\section{Conclusions}

We have demonstrated that in the ballistic regime the thermal boundary resistance between sintered metal and \textsuperscript{3}He is dominated by the collisions of quasiparticles with sinter walls. The probability of an eventually recombining collision can be calculated using the thermal time constant. The probability is surprisingly very low, on the order of one in a thousand. By considering the geometric surface area quasiparticles have available to scatter with, we find good agreement of Kapitza resistance with previous measurements. To increase the thermal boundary conductivity it is therefore important to increase the frontal surface area the particles can collide with in the experimental volume.

When designing cells for ultra-low temperature experiments, increasing the geometric surface in contact with the \textsuperscript{3}He is an important factor. Future designs should aim to increase the contact area, possibly by using ridge-like protrusions. 

It is also important to consider the magnetic properties of the surface the quasiparticles scatter with. Covering the surfaces with two layers of non-magnetic solid \textsuperscript{4}He doubles the thermal boundary resistance. This provides further evidence of a magnetic energy transfer channel between liquid \textsuperscript{3}He and metals to explain deviations in Kapitza resistance from theory. A well-developed theory explaining the Kapitza resistance at ultra-low temperatures should include the magnetic properties and future sinter designs should aim at utilizing magnetic impurities.

It would be beneficial to extend the scope of the measurements by increasing the temperature range and probing different phases. Unfortunately, the described bolometry method can only be applied in the ballistic regime of the B phase and cannot be used in the A phase. Therefore, we were unable to make a similar measurement in the A phase. The method of heat transfer from superfluid \textsuperscript{3}He-A to sinter at ultra-low temperatures is an interesting question for further work, since the propagation of quasiparticles is significantly different~\cite{Pickett1994}. One would assume it is unlikely that the result in the B phase would then apply, but earlier measurements of the A phase found an exponential dependence on temperature similar to that in the B phase between \SI{230}{\micro\kelvin} and \SI{450}{\micro\kelvin} at zero pressure \cite{Fisher1991}.

\begin{acknowledgments}
We acknowledge M.G. Ward and A. Stokes for their excellent technical support. This work was funded by UK EPSRC (grants No. EP/P024203/1 and No. EP/P022197/1) and EU H2020 European Microkelvin Platform (Grant Agreement 824109). S.A. acknowledges support from the Jenny and Antti Wihuri Foundation. 
\end{acknowledgments}

\nocite{*}


\begin{thebibliography}{30}%
\makeatletter
\providecommand \@ifxundefined [1]{%
 \@ifx{#1\undefined}
}%
\providecommand \@ifnum [1]{%
 \ifnum #1\expandafter \@firstoftwo
 \else \expandafter \@secondoftwo
 \fi
}%
\providecommand \@ifx [1]{%
 \ifx #1\expandafter \@firstoftwo
 \else \expandafter \@secondoftwo
 \fi
}%
\providecommand \natexlab [1]{#1}%
\providecommand \enquote  [1]{``#1''}%
\providecommand \bibnamefont  [1]{#1}%
\providecommand \bibfnamefont [1]{#1}%
\providecommand \citenamefont [1]{#1}%
\providecommand \href@noop [0]{\@secondoftwo}%
\providecommand \href [0]{\begingroup \@sanitize@url \@href}%
\providecommand \@href[1]{\@@startlink{#1}\@@href}%
\providecommand \@@href[1]{\endgroup#1\@@endlink}%
\providecommand \@sanitize@url [0]{\catcode `\\12\catcode `\$12\catcode
  `\&12\catcode `\#12\catcode `\^12\catcode `\_12\catcode `\%12\relax}%
\providecommand \@@startlink[1]{}%
\providecommand \@@endlink[0]{}%
\providecommand \url  [0]{\begingroup\@sanitize@url \@url }%
\providecommand \@url [1]{\endgroup\@href {#1}{\urlprefix }}%
\providecommand \urlprefix  [0]{URL }%
\providecommand \Eprint [0]{\href }%
\providecommand \doibase [0]{https://doi.org/}%
\providecommand \selectlanguage [0]{\@gobble}%
\providecommand \bibinfo  [0]{\@secondoftwo}%
\providecommand \bibfield  [0]{\@secondoftwo}%
\providecommand \translation [1]{[#1]}%
\providecommand \BibitemOpen [0]{}%
\providecommand \bibitemStop [0]{}%
\providecommand \bibitemNoStop [0]{.\EOS\space}%
\providecommand \EOS [0]{\spacefactor3000\relax}%
\providecommand \BibitemShut  [1]{\csname bibitem#1\endcsname}%
\let\auto@bib@innerbib\@empty
\bibitem [{\citenamefont {Nakayama}(1989)}]{Nakayama1989}%
\BibitemOpen
\bibfield  {author} {\bibinfo {author} {\bibfnamefont {T.}~\bibnamefont
		{Nakayama}},\ }\bibfield  {title} {\bibinfo {title} {Chapter 3: Kapitza
		thermal boundary resistance and interactions of helium quasiparticles with
		surfaces},\ }in\ \href {https://doi.org/10.1016/s0079-6417(08)60042-6} {\emph
	{\bibinfo {booktitle} {Progress in Low Temperature Physics}}}\ (\bibinfo
{publisher} {Elsevier},\ \bibinfo {year} {1989})\ pp.\ \bibinfo {pages}
{115--194}\BibitemShut {NoStop}%
\bibitem [{\citenamefont {Franco}\ \emph {et~al.}(1984)\citenamefont {Franco},
	\citenamefont {Bossy},\ and\ \citenamefont {Godfrin}}]{Franco1984}%
\BibitemOpen
\bibfield  {author} {\bibinfo {author} {\bibfnamefont {H.}~\bibnamefont
		{Franco}}, \bibinfo {author} {\bibfnamefont {J.}~\bibnamefont {Bossy}},\ and\
	\bibinfo {author} {\bibfnamefont {H.}~\bibnamefont {Godfrin}},\ }\bibfield
{title} {\bibinfo {title} {Properties of sintered silver powders and their
		application in heat exchangers at millikelvin temperatures},\ }\href
{https://doi.org/10.1016/0011-2275(84)90006-7} {\bibfield  {journal}
	{\bibinfo  {journal} {Cryogenics}\ }\textbf {\bibinfo {volume} {24}},\
	\bibinfo {pages} {477} (\bibinfo {year} {1984})}\BibitemShut {NoStop}%
\bibitem [{\citenamefont {Castelijns}\ \emph {et~al.}(1985)\citenamefont
	{Castelijns}, \citenamefont {Coates}, \citenamefont {Gu\'enault},
	\citenamefont {Mussett},\ and\ \citenamefont {Pickett}}]{Castelijns1985}%
\BibitemOpen
\bibfield  {author} {\bibinfo {author} {\bibfnamefont {C.~A.~M.}\
		\bibnamefont {Castelijns}}, \bibinfo {author} {\bibfnamefont {K.~F.}\
		\bibnamefont {Coates}}, \bibinfo {author} {\bibfnamefont {A.~M.}\
		\bibnamefont {Gu\'enault}}, \bibinfo {author} {\bibfnamefont {S.~G.}\
		\bibnamefont {Mussett}},\ and\ \bibinfo {author} {\bibfnamefont {G.~R.}\
		\bibnamefont {Pickett}},\ }\bibfield  {title} {\bibinfo {title} {Exponential
		temperature dependence of the thermal boundary conductance between sintered
		silver and $^{3}\mathrm{He}$-$\mathit{B}$ in the region of low normal fluid
		density},\ }\href {https://doi.org/10.1103/PhysRevLett.55.2021} {\bibfield
	{journal} {\bibinfo  {journal} {Phys. Rev. Lett.}\ }\textbf {\bibinfo
		{volume} {55}},\ \bibinfo {pages} {2021} (\bibinfo {year}
	{1985})}\BibitemShut {NoStop}%
\bibitem [{\citenamefont {Parpia}(1985)}]{Parpia1985}%
\BibitemOpen
\bibfield  {author} {\bibinfo {author} {\bibfnamefont {J.~M.}\ \bibnamefont
		{Parpia}},\ }\bibfield  {title} {\bibinfo {title} {Anomalous thermal boundary
		resistance of superfluid $^{3}\mathrm{He}$-$\mathit{B}$},\ }\href
{https://doi.org/10.1103/PhysRevB.32.7564} {\bibfield  {journal} {\bibinfo
		{journal} {Phys. Rev. B}\ }\textbf {\bibinfo {volume} {32}},\ \bibinfo
	{pages} {7564} (\bibinfo {year} {1985})}\BibitemShut {NoStop}%
\bibitem [{\citenamefont {Voncken}\ \emph {et~al.}(1996)\citenamefont
	{Voncken}, \citenamefont {Riese}, \citenamefont {Roobol}, \citenamefont
	{K\"onig},\ and\ \citenamefont {Pobell}}]{Voncken1996}%
\BibitemOpen
\bibfield  {author} {\bibinfo {author} {\bibfnamefont {A.~P.~J.}\
		\bibnamefont {Voncken}}, \bibinfo {author} {\bibfnamefont {D.}~\bibnamefont
		{Riese}}, \bibinfo {author} {\bibfnamefont {L.~P.}\ \bibnamefont {Roobol}},
	\bibinfo {author} {\bibfnamefont {R.}~\bibnamefont {K\"onig}},\ and\ \bibinfo
	{author} {\bibfnamefont {F.}~\bibnamefont {Pobell}},\ }\bibfield  {title}
{\bibinfo {title} {Thermal boundary resistance between liquid helium and
		silver sinter at low temperatures},\ }\href
{https://doi.org/10.1007/bf00754629} {\bibfield  {journal} {\bibinfo
		{journal} {Journal of Low Temperature Physics}\ }\textbf {\bibinfo {volume}
		{105}},\ \bibinfo {pages} {93} (\bibinfo {year} {1996})}\BibitemShut
{NoStop}%
\bibitem [{\citenamefont {Tholen}\ and\ \citenamefont
	{Parpia}(1993)}]{Tholen1993}%
\BibitemOpen
\bibfield  {author} {\bibinfo {author} {\bibfnamefont {S.~M.}\ \bibnamefont
		{Tholen}}\ and\ \bibinfo {author} {\bibfnamefont {J.~M.}\ \bibnamefont
		{Parpia}},\ }\bibfield  {title} {\bibinfo {title} {Effect of
		$^{4}\mathrm{He}$ on the surface scattering of $^{3}\mathrm{He}$},\ }\href
{https://doi.org/10.1103/PhysRevB.47.319} {\bibfield  {journal} {\bibinfo
		{journal} {Phys. Rev. B}\ }\textbf {\bibinfo {volume} {47}},\ \bibinfo
	{pages} {319} (\bibinfo {year} {1993})}\BibitemShut {NoStop}%
\bibitem [{\citenamefont {Levitin}\ \emph {et~al.}(2013)\citenamefont
	{Levitin}, \citenamefont {Bennett}, \citenamefont {Casey}, \citenamefont
	{Cowan}, \citenamefont {Saunders}, \citenamefont {Drung}, \citenamefont
	{Schurig},\ and\ \citenamefont {Parpia}}]{Levitin2013}%
\BibitemOpen
\bibfield  {author} {\bibinfo {author} {\bibfnamefont {L.~V.}\ \bibnamefont
		{Levitin}}, \bibinfo {author} {\bibfnamefont {R.~G.}\ \bibnamefont
		{Bennett}}, \bibinfo {author} {\bibfnamefont {A.}~\bibnamefont {Casey}},
	\bibinfo {author} {\bibfnamefont {B.}~\bibnamefont {Cowan}}, \bibinfo
	{author} {\bibfnamefont {J.}~\bibnamefont {Saunders}}, \bibinfo {author}
	{\bibfnamefont {D.}~\bibnamefont {Drung}}, \bibinfo {author} {\bibfnamefont
		{T.}~\bibnamefont {Schurig}},\ and\ \bibinfo {author} {\bibfnamefont {J.~M.}\
		\bibnamefont {Parpia}},\ }\bibfield  {title} {\bibinfo {title} {Phase diagram
		of the topological superfluid $^3${He} confined in a nanoscale slab
		geometry},\ }\href {https://doi.org/10.1126/science.1233621} {\bibfield
	{journal} {\bibinfo  {journal} {Science}\ }\textbf {\bibinfo {volume}
		{340}},\ \bibinfo {pages} {841} (\bibinfo {year} {2013})}\BibitemShut
{NoStop}%
\bibitem [{\citenamefont {K\"onig}\ \emph
	{et~al.}(1998{\natexlab{a}})\citenamefont {K\"onig}, \citenamefont
	{Herrmannsd\"orfer}, \citenamefont {Schindler},\ and\ \citenamefont
	{Usherov-Marshak}}]{Koenig1998}%
\BibitemOpen
\bibfield  {author} {\bibinfo {author} {\bibfnamefont {R.}~\bibnamefont
		{K\"onig}}, \bibinfo {author} {\bibfnamefont {T.}~\bibnamefont
		{Herrmannsd\"orfer}}, \bibinfo {author} {\bibfnamefont {A.}~\bibnamefont
		{Schindler}},\ and\ \bibinfo {author} {\bibfnamefont {I.}~\bibnamefont
		{Usherov-Marshak}},\ }\bibfield  {title} {\bibinfo {title} {The origin of the
		magnetic channel of the thermal boundary resistance between liquid $^3${He}
		and {Ag} sinters at very low temperatures},\ }\href
{https://doi.org/10.1023/a:1022587830162} {\bibfield  {journal} {\bibinfo
		{journal} {Journal of Low Temperature Physics}\ }\textbf {\bibinfo {volume}
		{113}},\ \bibinfo {pages} {969} (\bibinfo {year}
	{1998}{\natexlab{a}})}\BibitemShut {NoStop}%
\bibitem [{\citenamefont {K\"onig}\ \emph
	{et~al.}(1998{\natexlab{b}})\citenamefont {K\"onig}, \citenamefont
	{Herrmannsdörfer},\ and\ \citenamefont {Batko}}]{Koenig1998a}%
\BibitemOpen
\bibfield  {author} {\bibinfo {author} {\bibfnamefont {R.}~\bibnamefont
		{K\"onig}}, \bibinfo {author} {\bibfnamefont {T.}~\bibnamefont
		{Herrmannsdörfer}},\ and\ \bibinfo {author} {\bibfnamefont {I.}~\bibnamefont
		{Batko}},\ }\bibfield  {title} {\bibinfo {title} {Magnetization of {Ag}
		sinters made of compressed particles of submicron grain size and their
		coupling to liquid $^3${He}},\ }\href
{https://doi.org/10.1103/physrevlett.80.4787} {\bibfield  {journal} {\bibinfo
		{journal} {Physical Review Letters}\ }\textbf {\bibinfo {volume} {80}},\
	\bibinfo {pages} {4787} (\bibinfo {year} {1998}{\natexlab{b}})}\BibitemShut
{NoStop}%
\bibitem [{\citenamefont {Dmitriev}\ \emph {et~al.}(2015)\citenamefont
	{Dmitriev}, \citenamefont {Senin}, \citenamefont {Soldatov},\ and\
	\citenamefont {Yudin}}]{Dmitriev2015}%
\BibitemOpen
\bibfield  {author} {\bibinfo {author} {\bibfnamefont {V.~V.}\ \bibnamefont
		{Dmitriev}}, \bibinfo {author} {\bibfnamefont {A.~A.}\ \bibnamefont {Senin}},
	\bibinfo {author} {\bibfnamefont {A.~A.}\ \bibnamefont {Soldatov}},\ and\
	\bibinfo {author} {\bibfnamefont {A.~N.}\ \bibnamefont {Yudin}},\ }\bibfield
{title} {\bibinfo {title} {Polar phase of superfluid $^{3}\mathrm{He}$ in
		anisotropic aerogel},\ }\href
{https://doi.org/10.1103/PhysRevLett.115.165304} {\bibfield  {journal}
	{\bibinfo  {journal} {Phys. Rev. Lett.}\ }\textbf {\bibinfo {volume} {115}},\
	\bibinfo {pages} {165304} (\bibinfo {year} {2015})}\BibitemShut {NoStop}%
\bibitem [{\citenamefont {Zimmerman}\ \emph {et~al.}(2020)\citenamefont
	{Zimmerman}, \citenamefont {Nguyen}, \citenamefont {Scott},\ and\
	\citenamefont {Halperin}}]{Zimmerman2020}%
\BibitemOpen
\bibfield  {author} {\bibinfo {author} {\bibfnamefont {A.}~\bibnamefont
		{Zimmerman}}, \bibinfo {author} {\bibfnamefont {M.}~\bibnamefont {Nguyen}},
	\bibinfo {author} {\bibfnamefont {J.}~\bibnamefont {Scott}},\ and\ \bibinfo
	{author} {\bibfnamefont {W.}~\bibnamefont {Halperin}},\ }\bibfield  {title}
{\bibinfo {title} {Effect of magnetic impurities on superfluid $^3${He}},\
}\href {https://doi.org/10.1103/physrevlett.124.025302} {\bibfield  {journal}
	{\bibinfo  {journal} {Physical Review Letters}\ }\textbf {\bibinfo {volume}
		{124}},\ \bibinfo {pages} {025302} (\bibinfo {year} {2020})}\BibitemShut
{NoStop}%
\bibitem [{\citenamefont {Heikkinen}\ \emph {et~al.}()\citenamefont
	{Heikkinen}, \citenamefont {Casey}, \citenamefont {Levitin}, \citenamefont
	{Rojas}, \citenamefont {Vorontsov}, \citenamefont {Sharma}, \citenamefont
	{Zhelev}, \citenamefont {Parpia},\ and\ \citenamefont
	{Saunders}}]{Heikkinen2019}%
\BibitemOpen
\bibfield  {author} {\bibinfo {author} {\bibfnamefont {P.~J.}\ \bibnamefont
		{Heikkinen}}, \bibinfo {author} {\bibfnamefont {A.}~\bibnamefont {Casey}},
	\bibinfo {author} {\bibfnamefont {L.~V.}\ \bibnamefont {Levitin}}, \bibinfo
	{author} {\bibfnamefont {X.}~\bibnamefont {Rojas}}, \bibinfo {author}
	{\bibfnamefont {A.}~\bibnamefont {Vorontsov}}, \bibinfo {author}
	{\bibfnamefont {P.}~\bibnamefont {Sharma}}, \bibinfo {author} {\bibfnamefont
		{N.}~\bibnamefont {Zhelev}}, \bibinfo {author} {\bibfnamefont {J.~M.}\
		\bibnamefont {Parpia}},\ and\ \bibinfo {author} {\bibfnamefont
		{J.}~\bibnamefont {Saunders}},\ }\bibfield  {title} {\bibinfo {title}
	{Fragility of surface states in topological superfluid $^3${He}},\ }\href
{https://arxiv.org/abs/1909.04210} {\ }\Eprint
{https://arxiv.org/abs/arXiv:1909.04210} {arXiv:1909.04210} \BibitemShut
{NoStop}%
\bibitem [{\citenamefont {Dmitriev}\ \emph {et~al.}(2018)\citenamefont
	{Dmitriev}, \citenamefont {Soldatov},\ and\ \citenamefont
	{Yudin}}]{Dmitriev2018}%
\BibitemOpen
\bibfield  {author} {\bibinfo {author} {\bibfnamefont {V.}~\bibnamefont
		{Dmitriev}}, \bibinfo {author} {\bibfnamefont {A.}~\bibnamefont {Soldatov}},\
	and\ \bibinfo {author} {\bibfnamefont {A.}~\bibnamefont {Yudin}},\ }\bibfield
{title} {\bibinfo {title} {Effect of magnetic boundary conditions on
		superfluid $^3${He} in nematic aerogel},\ }\href
{https://doi.org/10.1103/physrevlett.120.075301} {\bibfield  {journal}
	{\bibinfo  {journal} {Physical Review Letters}\ }\textbf {\bibinfo {volume}
		{120}},\ \bibinfo {pages} {075301} (\bibinfo {year} {2018})}\BibitemShut
{NoStop}%
\bibitem [{\citenamefont {Carney}\ \emph {et~al.}(1989)\citenamefont {Carney},
	\citenamefont {Coates}, \citenamefont {Gu\'enault}, \citenamefont {Pickett},\
	and\ \citenamefont {Spencer}}]{Carney1989}%
\BibitemOpen
\bibfield  {author} {\bibinfo {author} {\bibfnamefont {J.~P.}\ \bibnamefont
		{Carney}}, \bibinfo {author} {\bibfnamefont {K.~F.}\ \bibnamefont {Coates}},
	\bibinfo {author} {\bibfnamefont {A.~M.}\ \bibnamefont {Gu\'enault}},
	\bibinfo {author} {\bibfnamefont {G.~R.}\ \bibnamefont {Pickett}},\ and\
	\bibinfo {author} {\bibfnamefont {G.~F.}\ \bibnamefont {Spencer}},\
}\bibfield  {title} {\bibinfo {title} {Thermal behavior of quasiparticles in
		$^3${He-B}},\ }\href {https://doi.org/10.1007/bf00681737} {\bibfield
	{journal} {\bibinfo  {journal} {Journal of Low Temperature Physics}\ }\textbf
	{\bibinfo {volume} {76}},\ \bibinfo {pages} {417} (\bibinfo {year}
	{1989})}\BibitemShut {NoStop}%
\bibitem [{\citenamefont {Cousins}\ \emph {et~al.}(1994)\citenamefont
	{Cousins}, \citenamefont {Fisher}, \citenamefont {Gu{\'{e}}nault},
	\citenamefont {Pickett}, \citenamefont {Smith},\ and\ \citenamefont
	{Turner}}]{Cousins1994}%
\BibitemOpen
\bibfield  {author} {\bibinfo {author} {\bibfnamefont {D.~J.}\ \bibnamefont
		{Cousins}}, \bibinfo {author} {\bibfnamefont {S.~N.}\ \bibnamefont {Fisher}},
	\bibinfo {author} {\bibfnamefont {A.~M.}\ \bibnamefont {Gu{\'{e}}nault}},
	\bibinfo {author} {\bibfnamefont {G.~R.}\ \bibnamefont {Pickett}}, \bibinfo
	{author} {\bibfnamefont {E.~N.}\ \bibnamefont {Smith}},\ and\ \bibinfo
	{author} {\bibfnamefont {R.~P.}\ \bibnamefont {Turner}},\ }\bibfield  {title}
{\bibinfo {title} {T$^{-3}$ temperature dependence and a length scale for the
		thermal boundary resistance between a saturated dilute $^3${He}-$^4${He}
		solution and sintered silver},\ }\href
{https://doi.org/10.1103/physrevlett.73.2583} {\bibfield  {journal} {\bibinfo
		{journal} {Physical Review Letters}\ }\textbf {\bibinfo {volume} {73}},\
	\bibinfo {pages} {2583} (\bibinfo {year} {1994})}\BibitemShut {NoStop}%
\bibitem [{\citenamefont {Keith}\ and\ \citenamefont {Ward}(1984)}]{Keith1984}%
\BibitemOpen
\bibfield  {author} {\bibinfo {author} {\bibfnamefont {V.}~\bibnamefont
		{Keith}}\ and\ \bibinfo {author} {\bibfnamefont {M.}~\bibnamefont {Ward}},\
}\bibfield  {title} {\bibinfo {title} {A recipe for sintering submicron
		silver powders},\ }\href {https://doi.org/10.1016/0011-2275(84)90151-6}
{\bibfield  {journal} {\bibinfo  {journal} {Cryogenics}\ }\textbf {\bibinfo
		{volume} {24}},\ \bibinfo {pages} {249} (\bibinfo {year} {1984})}\BibitemShut
{NoStop}%
\bibitem [{\citenamefont {Autti}\ \emph {et~al.}()\citenamefont {Autti},
	\citenamefont {Ahlstrom},  \citenamefont {Haley}, \citenamefont {Jennings},
	\citenamefont {Pickett},  \citenamefont {Schanen}, \citenamefont {Soldatov},
	\citenamefont {Tsepelin},  \citenamefont {Vonka}, \citenamefont {Wilcox},
	\citenamefont {Woods},\ and\ \citenamefont
	{Zmeev}}]{Autti2020}%
\BibitemOpen
\bibfield  {author} {\bibinfo {author} {\bibfnamefont {S.}~\bibnamefont
		{Autti}}, \bibinfo {author} {\bibfnamefont {S.~L.} \bibnamefont
		{Ahlstrom}}  \bibinfo {author} {\bibfnamefont {R.~P.}\ \bibnamefont {Haley}},
	\bibinfo {author} {\bibfnamefont {A.}~\bibnamefont {Jennings}}, \bibinfo
	{author} {\bibfnamefont {G.~R.}\ \bibnamefont {Pickett}}, \bibinfo {author}
	{\bibfnamefont {R.}~\bibnamefont {Schanen}}, \bibinfo {author} {\bibfnamefont
		{A.~A.}\ \bibnamefont {Soldatov}}, \bibinfo {author} {\bibfnamefont
		{V.}~\bibnamefont {Tsepelin}}, \bibinfo {author} {\bibfnamefont
		{J.}~\bibnamefont {Vonka}}, \bibinfo {author} {\bibfnamefont
		{T.}~\bibnamefont {Wilcox}}, \bibinfo {author} {\bibfnamefont {A.~J.} \bibnamefont
		{Woods}}\ and\ \bibinfo {author} {\bibfnamefont {D.~E.}\
		\bibnamefont {Zmeev}},\ }\bibfield  {title} {\bibinfo {title} {Fundamental
		dissipation due to bound fermions in the zero-temperature limit},\ }\href
{https://doi.org/10.1038/s41467-020-18499-1}{\bibfield  {journal} {\bibinfo
		{journal} {Nature Communications}\ }\textbf {\bibinfo {volume} {11}},\ \bibinfo
	{pages} {4742} (\bibinfo {year} {2020})}\BibitemShut {NoStop}%
\bibitem [{\citenamefont {Gu\'{e}nault}\ \emph {et~al.}(1986)\citenamefont
	{Gu\'{e}nault}, \citenamefont {Keith}, \citenamefont {Kennedy}, \citenamefont
	{Mussett},\ and\ \citenamefont {Pickett}}]{Guenault1986}%
\BibitemOpen
\bibfield  {author} {\bibinfo {author} {\bibfnamefont {A.~M.}\ \bibnamefont
		{Gu\'{e}nault}}, \bibinfo {author} {\bibfnamefont {V.}~\bibnamefont {Keith}},
	\bibinfo {author} {\bibfnamefont {C.~J.}\ \bibnamefont {Kennedy}}, \bibinfo
	{author} {\bibfnamefont {S.~G.}\ \bibnamefont {Mussett}},\ and\ \bibinfo
	{author} {\bibfnamefont {G.~R.}\ \bibnamefont {Pickett}},\ }\bibfield
{title} {\bibinfo {title} {The mechanical behavior of a vibrating wire in
		superfluid $^3${He}-$\mathit{B}$ in the ballistic limit},\ }\href
{https://doi.org/10.1007/bf00683408} {\bibfield  {journal} {\bibinfo
		{journal} {Journal of Low Temperature Physics}\ }\textbf {\bibinfo {volume}
		{62}},\ \bibinfo {pages} {511} (\bibinfo {year} {1986})}\BibitemShut
{NoStop}%
\bibitem [{\citenamefont {Zmeev}(2014)}]{Zmeev2013}%
\BibitemOpen
\bibfield  {author} {\bibinfo {author} {\bibfnamefont {D.~E.}\ \bibnamefont
		{Zmeev}},\ }\bibfield  {title} {\bibinfo {title} {A method for driving an
		oscillator at a quasi-uniform velocity},\ }\href
{https://doi.org/10.1007/s10909-013-0942-2} {\bibfield  {journal} {\bibinfo
		{journal} {Journal of Low Temperature Physics}\ }\textbf {\bibinfo {volume}
		{175}},\ \bibinfo {pages} {480} (\bibinfo {year} {2014})}\BibitemShut
{NoStop}%
\bibitem [{\citenamefont {Bradley}\ \emph {et~al.}(2016)\citenamefont
	{Bradley}, \citenamefont {Fisher}, \citenamefont {Gu\'{e}nault},
	\citenamefont {Haley}, \citenamefont {Lawson}, \citenamefont {Pickett},
	\citenamefont {Schanen}, \citenamefont {Skyba}, \citenamefont {Tsepelin},\
	and\ \citenamefont {Zmeev}}]{Bradley2016}%
\BibitemOpen
\bibfield  {author} {\bibinfo {author} {\bibfnamefont {D.~I.}\ \bibnamefont
		{Bradley}}, \bibinfo {author} {\bibfnamefont {S.~N.}\ \bibnamefont {Fisher}},
	\bibinfo {author} {\bibfnamefont {A.~M.}\ \bibnamefont {Gu\'{e}nault}},
	\bibinfo {author} {\bibfnamefont {R.~P.}\ \bibnamefont {Haley}}, \bibinfo
	{author} {\bibfnamefont {C.~R.}\ \bibnamefont {Lawson}}, \bibinfo {author}
	{\bibfnamefont {G.~R.}\ \bibnamefont {Pickett}}, \bibinfo {author}
	{\bibfnamefont {R.}~\bibnamefont {Schanen}}, \bibinfo {author} {\bibfnamefont
		{M.}~\bibnamefont {Skyba}}, \bibinfo {author} {\bibfnamefont
		{V.}~\bibnamefont {Tsepelin}},\ and\ \bibinfo {author} {\bibfnamefont
		{D.~E.}\ \bibnamefont {Zmeev}},\ }\bibfield  {title} {\bibinfo {title}
	{Breaking the superfluid speed limit in a fermionic condensate},\ }\href
{http://dx.doi.org/10.1038/nphys3813} {\bibfield  {journal} {\bibinfo
		{journal} {Nature Physics}\ }\textbf {\bibinfo {volume} {12}},\ \bibinfo
	{pages} {1017} (\bibinfo {year} {2016})}\BibitemShut {NoStop}%
\bibitem [{\citenamefont {Winkelmann}\ \emph {et~al.}(2007)\citenamefont
	{Winkelmann}, \citenamefont {Elbs}, \citenamefont {Bunkov}, \citenamefont
	{Collin}, \citenamefont {Godfrin},\ and\ \citenamefont
	{Krusius}}]{Winkelmann2007}%
\BibitemOpen
\bibfield  {author} {\bibinfo {author} {\bibfnamefont {C.}~\bibnamefont
		{Winkelmann}}, \bibinfo {author} {\bibfnamefont {J.}~\bibnamefont {Elbs}},
	\bibinfo {author} {\bibfnamefont {Y.~M.}\ \bibnamefont {Bunkov}}, \bibinfo
	{author} {\bibfnamefont {E.}~\bibnamefont {Collin}}, \bibinfo {author}
	{\bibfnamefont {H.}~\bibnamefont {Godfrin}},\ and\ \bibinfo {author}
	{\bibfnamefont {M.}~\bibnamefont {Krusius}},\ }\bibfield  {title} {\bibinfo
	{title} {Bolometric calibration of a superfluid $^3${He} detector for dark
		matter search: Direct measurement of the scintillated energy fraction for
		neutron, electron and muon events},\ }\href
{https://doi.org/10.1016/j.nima.2007.01.180} {\bibfield  {journal} {\bibinfo
		{journal} {Nuclear Instruments and Methods in Physics Research Section A:
			Accelerators, Spectrometers, Detectors and Associated Equipment}\ }\textbf
	{\bibinfo {volume} {574}},\ \bibinfo {pages} {264} (\bibinfo {year}
	{2007})}\BibitemShut {NoStop}%
\bibitem [{\citenamefont {Ba\"uerle}\ \emph {et~al.}(1998)\citenamefont
	{Ba\"uerle}, \citenamefont {Bunkov}, \citenamefont {Fisher},\ and\
	\citenamefont {Godfrin}}]{Baeuerle1998}%
\BibitemOpen
\bibfield  {author} {\bibinfo {author} {\bibfnamefont {C.}~\bibnamefont
		{Ba\"uerle}}, \bibinfo {author} {\bibfnamefont {Y.~M.}\ \bibnamefont
		{Bunkov}}, \bibinfo {author} {\bibfnamefont {S.~N.}\ \bibnamefont {Fisher}},\
	and\ \bibinfo {author} {\bibfnamefont {H.}~\bibnamefont {Godfrin}},\
}\bibfield  {title} {\bibinfo {title} {Temperature scale and heat capacity of
		superfluid $^3${He-B} in the 100~$\mu$k range},\ }\href
{https://doi.org/10.1103/physrevb.57.14381} {\bibfield  {journal} {\bibinfo
		{journal} {Physical Review B}\ }\textbf {\bibinfo {volume} {57}},\ \bibinfo
	{pages} {14381} (\bibinfo {year} {1998})}\BibitemShut {NoStop}%
\bibitem [{\citenamefont {Elbs}\ \emph {et~al.}(2007)\citenamefont {Elbs},
	\citenamefont {Winkelmann}, \citenamefont {Bunkov}, \citenamefont {Collin},\
	and\ \citenamefont {Godfrin}}]{Elbs2007}%
\BibitemOpen
\bibfield  {author} {\bibinfo {author} {\bibfnamefont {J.}~\bibnamefont
		{Elbs}}, \bibinfo {author} {\bibfnamefont {C.}~\bibnamefont {Winkelmann}},
	\bibinfo {author} {\bibfnamefont {Y.~M.}\ \bibnamefont {Bunkov}}, \bibinfo
	{author} {\bibfnamefont {E.}~\bibnamefont {Collin}},\ and\ \bibinfo {author}
	{\bibfnamefont {H.}~\bibnamefont {Godfrin}},\ }\bibfield  {title} {\bibinfo
	{title} {Heat capacity of adsorbed helium-3 at ultra-low temperatures},\
}\href {https://doi.org/10.1007/s10909-007-9428-4} {\bibfield  {journal}
	{\bibinfo  {journal} {Journal of Low Temperature Physics}\ }\textbf {\bibinfo
		{volume} {148}},\ \bibinfo {pages} {749} (\bibinfo {year}
	{2007})}\BibitemShut {NoStop}%
\bibitem [{\citenamefont {Serene}\ and\ \citenamefont
	{Rainer}(1983)}]{Serene1983}%
\BibitemOpen
\bibfield  {author} {\bibinfo {author} {\bibfnamefont {J.}~\bibnamefont
		{Serene}}\ and\ \bibinfo {author} {\bibfnamefont {D.}~\bibnamefont
		{Rainer}},\ }\bibfield  {title} {\bibinfo {title} {The quasiclassical
		approach to superfluid \textsuperscript{3}{He}},\ }\href
{https://doi.org/10.1016/0370-1573(83)90051-0} {\bibfield  {journal}
	{\bibinfo  {journal} {Physics Reports}\ }\textbf {\bibinfo {volume} {101}},\
	\bibinfo {pages} {221} (\bibinfo {year} {1983})}\BibitemShut {NoStop}%
\bibitem [{\citenamefont {Hall}\ \emph {et~al.}(1992)\citenamefont {Hall},
	\citenamefont {Tholen}, \citenamefont {Lane}, \citenamefont {Kotsubo},\ and\
	\citenamefont {Parpia}}]{Hall1992}%
\BibitemOpen
\bibfield  {author} {\bibinfo {author} {\bibfnamefont {T.}~\bibnamefont
		{Hall}}, \bibinfo {author} {\bibfnamefont {S.~M.}\ \bibnamefont {Tholen}},
	\bibinfo {author} {\bibfnamefont {K.~R.}\ \bibnamefont {Lane}}, \bibinfo
	{author} {\bibfnamefont {V.}~\bibnamefont {Kotsubo}},\ and\ \bibinfo {author}
	{\bibfnamefont {J.~M.}\ \bibnamefont {Parpia}},\ }\bibfield  {title}
{\bibinfo {title} {The superfluid fraction of $^3${He} confined in pores of
		sintered silver},\ }\href {https://doi.org/10.1007/bf00683893} {\bibfield
	{journal} {\bibinfo  {journal} {Journal of Low Temperature Physics}\ }\textbf
	{\bibinfo {volume} {89}},\ \bibinfo {pages} {897} (\bibinfo {year}
	{1992})}\BibitemShut {NoStop}%
\bibitem [{\citenamefont {Osheroff}\ and\ \citenamefont
	{Richardson}(1985)}]{Osheroff1985}%
\BibitemOpen
\bibfield  {author} {\bibinfo {author} {\bibfnamefont {D.~D.}\ \bibnamefont
		{Osheroff}}\ and\ \bibinfo {author} {\bibfnamefont {R.~C.}\ \bibnamefont
		{Richardson}},\ }\bibfield  {title} {\bibinfo {title} {Novel magnetic field
		dependence of the coupling of excitations between two fermion fluids},\
}\href {https://doi.org/10.1103/PhysRevLett.54.1178} {\bibfield  {journal}
	{\bibinfo  {journal} {Phys. Rev. Lett.}\ }\textbf {\bibinfo {volume} {54}},\
	\bibinfo {pages} {1178} (\bibinfo {year} {1985})}\BibitemShut {NoStop}%
\bibitem [{\citenamefont {Avenel}\ \emph {et~al.}(1973)\citenamefont {Avenel},
	\citenamefont {Berglund}, \citenamefont {Gylling}, \citenamefont {Phillips},
	\citenamefont {Vetleseter},\ and\ \citenamefont {Vuorio}}]{Avenel1973}%
\BibitemOpen
\bibfield  {author} {\bibinfo {author} {\bibfnamefont {O.}~\bibnamefont
		{Avenel}}, \bibinfo {author} {\bibfnamefont {M.~P.}\ \bibnamefont
		{Berglund}}, \bibinfo {author} {\bibfnamefont {R.~G.}\ \bibnamefont
		{Gylling}}, \bibinfo {author} {\bibfnamefont {N.~E.}\ \bibnamefont
		{Phillips}}, \bibinfo {author} {\bibfnamefont {A.}~\bibnamefont
		{Vetleseter}},\ and\ \bibinfo {author} {\bibfnamefont {M.}~\bibnamefont
		{Vuorio}},\ }\bibfield  {title} {\bibinfo {title} {Improved thermal contact
		at ultralow temperatures between $^{3}\mathrm{He}$ and metals containing
		magnetic impurities},\ }\href {https://doi.org/10.1103/PhysRevLett.31.76}
{\bibfield  {journal} {\bibinfo  {journal} {Phys. Rev. Lett.}\ }\textbf
	{\bibinfo {volume} {31}},\ \bibinfo {pages} {76} (\bibinfo {year}
	{1973})}\BibitemShut {NoStop}%
\bibitem [{\citenamefont {K\"onig}\ \emph {et~al.}(1997)\citenamefont
	{K\"onig}, \citenamefont {Herrmannsd\"orfer}, \citenamefont {Riese},\ and\
	\citenamefont {Jansen}}]{Koenig1997}%
\BibitemOpen
\bibfield  {author} {\bibinfo {author} {\bibfnamefont {R.}~\bibnamefont
		{K\"onig}}, \bibinfo {author} {\bibfnamefont {T.}~\bibnamefont
		{Herrmannsd\"orfer}}, \bibinfo {author} {\bibfnamefont {D.}~\bibnamefont
		{Riese}},\ and\ \bibinfo {author} {\bibfnamefont {W.}~\bibnamefont
		{Jansen}},\ }\bibfield  {title} {\bibinfo {title} {Magnetic properties of
		{Ag} sinters and their possible impact on the coupling to liquid $^3${He} at
		very low temperatures},\ }\href {https://doi.org/10.1007/bf02395924}
{\bibfield  {journal} {\bibinfo  {journal} {Journal of Low Temperature
			Physics}\ }\textbf {\bibinfo {volume} {106}},\ \bibinfo {pages} {581}
	(\bibinfo {year} {1997})}\BibitemShut {NoStop}%
\bibitem [{\citenamefont {Pickett}\ \emph {et~al.}(1994)\citenamefont
	{Pickett}, \citenamefont {Enrico}, \citenamefont {Fisher}, \citenamefont
	{Gu{\'{e}}nault},\ and\ \citenamefont {Torizuka}}]{Pickett1994}%
\BibitemOpen
\bibfield  {author} {\bibinfo {author} {\bibfnamefont {G.}~\bibnamefont
		{Pickett}}, \bibinfo {author} {\bibfnamefont {M.}~\bibnamefont {Enrico}},
	\bibinfo {author} {\bibfnamefont {S.}~\bibnamefont {Fisher}}, \bibinfo
	{author} {\bibfnamefont {A.}~\bibnamefont {Gu{\'{e}}nault}},\ and\ \bibinfo
	{author} {\bibfnamefont {K.}~\bibnamefont {Torizuka}},\ }\bibfield  {title}
{\bibinfo {title} {Superfluid $^3${He} at very low temperatures: a very
		unusual excitation gas},\ }\href
{https://doi.org/10.1016/0921-4526(94)90236-4} {\bibfield  {journal}
	{\bibinfo  {journal} {Physica B: Condensed Matter}\ }\textbf {\bibinfo
		{volume} {197}},\ \bibinfo {pages} {390} (\bibinfo {year}
	{1994})}\BibitemShut {NoStop}%
\bibitem [{\citenamefont {Fisher}\ \emph {et~al.}(1991)\citenamefont {Fisher},
	\citenamefont {Gu{\'{e}}nault}, \citenamefont {Jones}, \citenamefont
	{Kennedy},\ and\ \citenamefont {Pickett}}]{Fisher1991}%
\BibitemOpen
\bibfield  {author} {\bibinfo {author} {\bibfnamefont {S.~N.}\ \bibnamefont
		{Fisher}}, \bibinfo {author} {\bibfnamefont {A.~M.}\ \bibnamefont
		{Gu{\'{e}}nault}}, \bibinfo {author} {\bibfnamefont {A.~H.}\ \bibnamefont
		{Jones}}, \bibinfo {author} {\bibfnamefont {C.~J.}\ \bibnamefont {Kennedy}},\
	and\ \bibinfo {author} {\bibfnamefont {G.~R.}\ \bibnamefont {Pickett}},\
}\bibfield  {title} {\bibinfo {title} {Exponential boundary resistance
		between superfluid $^3${He-A} and silver sinter at temperatures down to 230\:
		$\mu$K},\ }\href {https://doi.org/10.1209/0295-5075/16/4/012} {\bibfield
	{journal} {\bibinfo  {journal} {Europhysics Letters ({EPL})}\ }\textbf
	{\bibinfo {volume} {16}},\ \bibinfo {pages} {385} (\bibinfo {year}
	{1991})}\BibitemShut {NoStop}%
\end{thebibliography}
\end{document}